\documentclass{appolb}
\usepackage{graphicx}
\usepackage{amsmath}
\usepackage{hyperref}
\graphicspath{{./figs/}}

\usepackage[square,numbers]{natbib}

\begin{document}

\title{\textsc{Graniitti}: towards a deep learning-enhanced Monte Carlo event generator for high-energy diffraction}
\author{Mikael Mieskolainen
\thanks{Presented at ``Diffraction and Low-$x$ 2022'', Corigliano Calabro (Italy), September 24-30, 2022.}
\address{Imperial College, Department of Physics, Prince Consort Road SW7, London}
}
\maketitle

\begin{abstract}
We introduce \textsc{Graniitti}, a new Monte Carlo event generator designed especially to solve the enigma of glueballs at the LHC. We discuss the available physics processes, compare the simulations against STAR data from RHIC and span ambitious future directions towards the first diffractive event generator with a deep learning-enhanced computational engine.
\end{abstract}

\section{Introduction}
Central-exclusive high-energy proton-proton processes probe in a unique way the non-perturbative structure and dynamics of quantum chromodynamics (QCD) via coherent diffractive exchanges. One of the most enigmatic questions on this topic is the nature of glueballs, the non-Abelian bound states of gluons. Since their prediction shortly after discovery of QCD, no strong experimental evidence has been found. It is expected is that their experimental discovery requires paying attention to special `glue rich' production processes such as diffraction and dedicated measurements simultaneously in several hadronic decay channels.

Current experiments at the LHC and RHIC are well suited to answer this question in the low-mass domain of fully exclusive (with forward protons) and semi-exclusive (forward protons dissociated) double-Pomeron-exchange mediated scattering processes. Interpreting the finite acceptance fiducial cross-section measurements requires a Monte Carlo event generator capable of simulating the required dynamical structure with exact kinematics. Especially the spin-parity structure of produced resonances and correlations between the forward protons and central decay products are crucial. \textsc{Graniitti} is the first public Monte Carlo event generator~\cite{Mieskolainen:2019jpv} which addresses these issues.

In Section~\ref{sec:processes} we briefly introduce the physics processes which can be simulated with \textsc{Graniitti}, in Section~\ref{sec:data} we compare simulations against recent differential cross-section measurements done in the STAR experiment at RHIC and conclude in Section~\ref{sec:technology} with future directions in terms of novel deep learning technology for high-energy diffraction.

\section{Physics processes}
\label{sec:processes}

\textsc{Graniitti}\footnote{Available at \href{https://github.com/mieskolainen/graniitti}{github.com/mieskolainen/graniitti} (MIT and GPLv3 license)} is designed bottom-up for $pp \rightarrow p^{(*)} + X + p^{(*)}$ processes, especially for the Regge domain where the scattering momentum transfer invariants $|t_1|, |t_2|$ are much smaller than center-of-mass energy squared $s$. Special emphasis is put on the non-perturbative production of a low-mass system $X$, where glueballs are expected. The current version supports proton-proton initial states. However, the simulator architecture easily generalizes to hadron-ion and lepton-hadron processes in the future. The following main soft process amplitudes are available:

\begin{figure}
\centering
\includegraphics[scale=0.29]{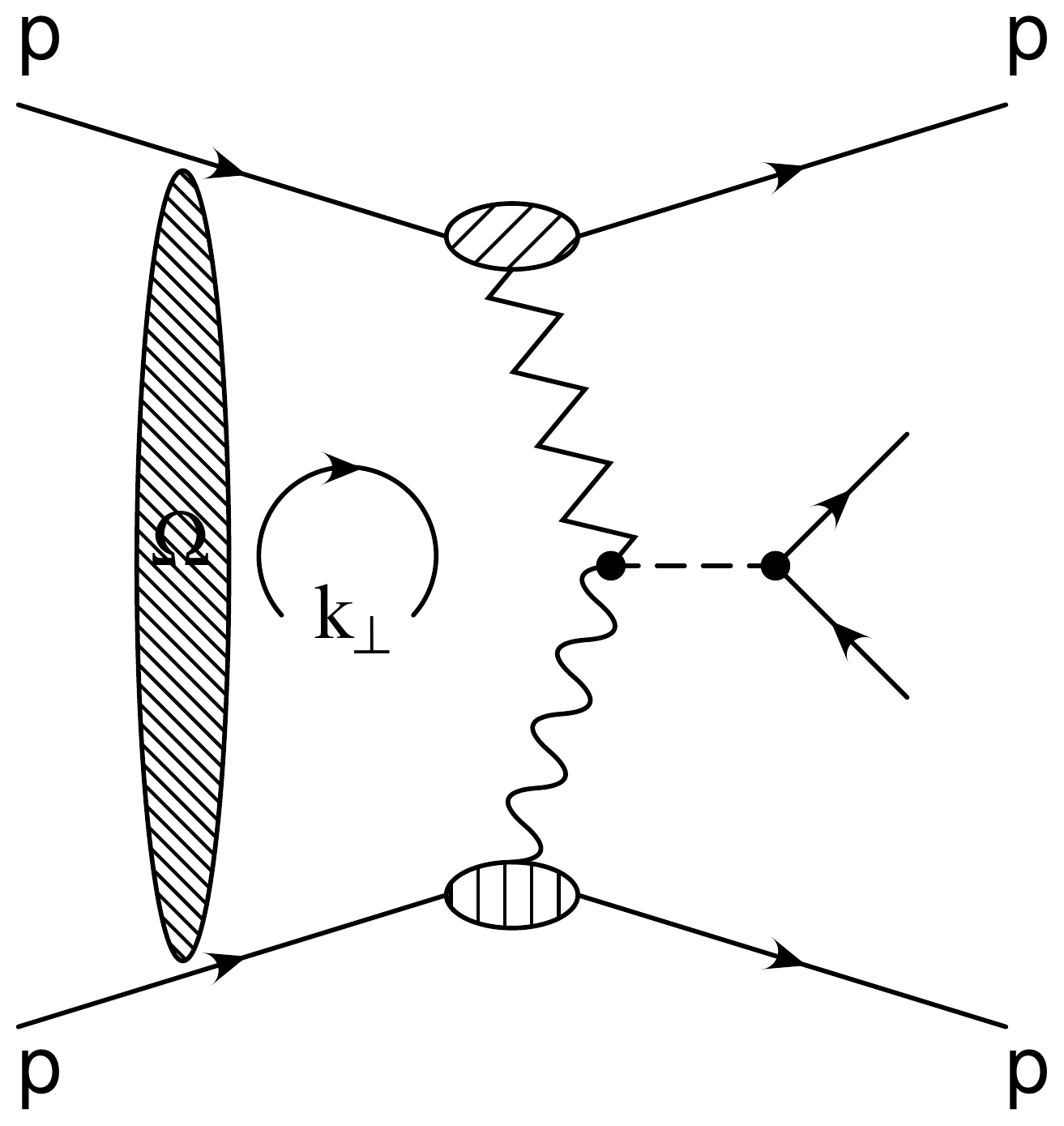}
\hspace{0.4em}
\includegraphics[scale=0.29]{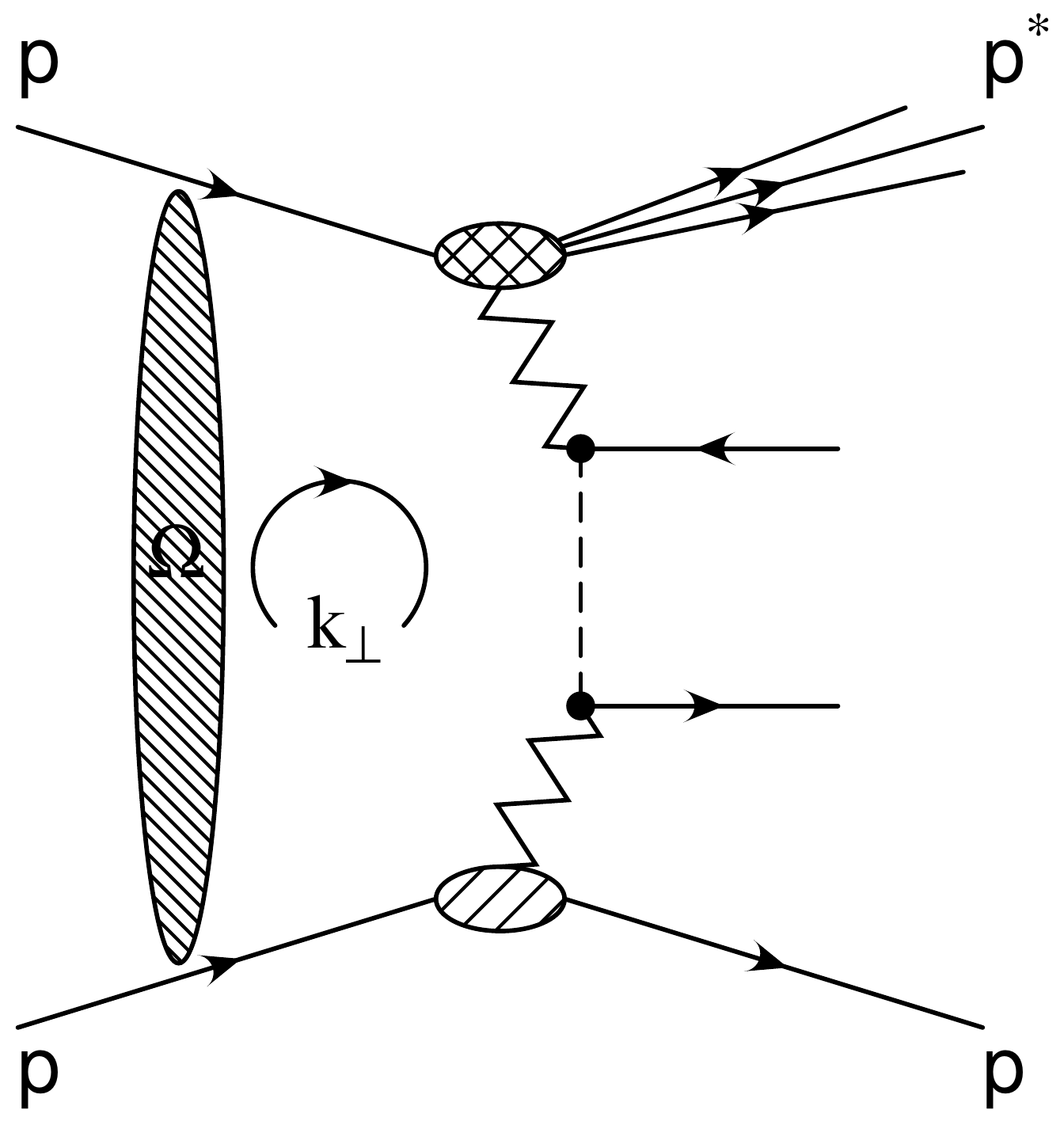}
\hspace{0.4em}
\includegraphics[scale=0.29]{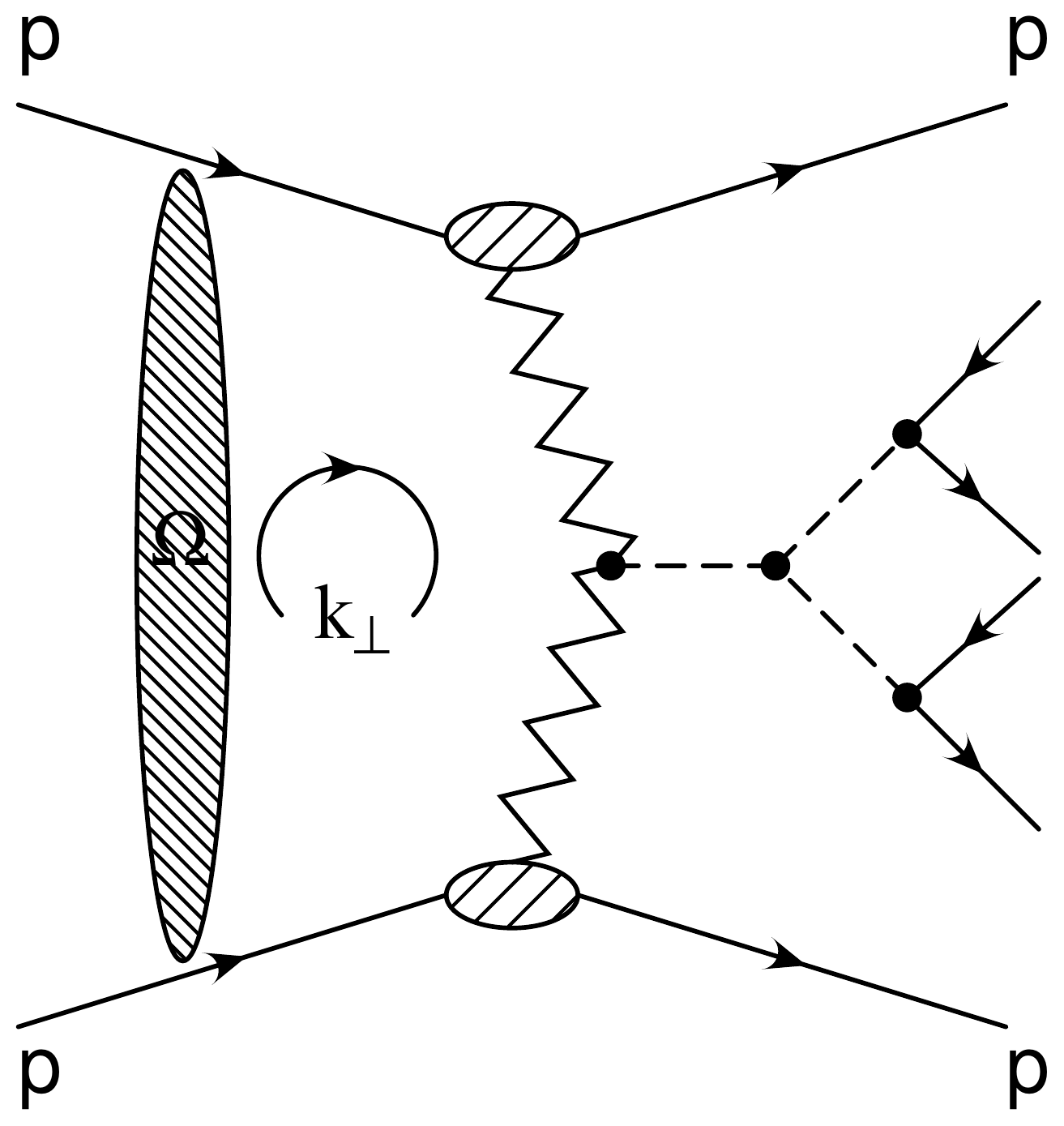}
\caption{Pomeron-Gamma-Resonance, Pomeron-Pomeron $\rightarrow$ central hadron pair + one excited forward proton $p^{*}$ and Pomeron-Pomeron-Resonance to a four-body central state. $k_T$ is the screening loop $2D$-momentum. For details, see~\cite{Mieskolainen:2019jpv}.}
\label{fig:diagrams}
\end{figure}
\vspace{1em}

\begin{figure}
\centering
\includegraphics[scale=0.3]{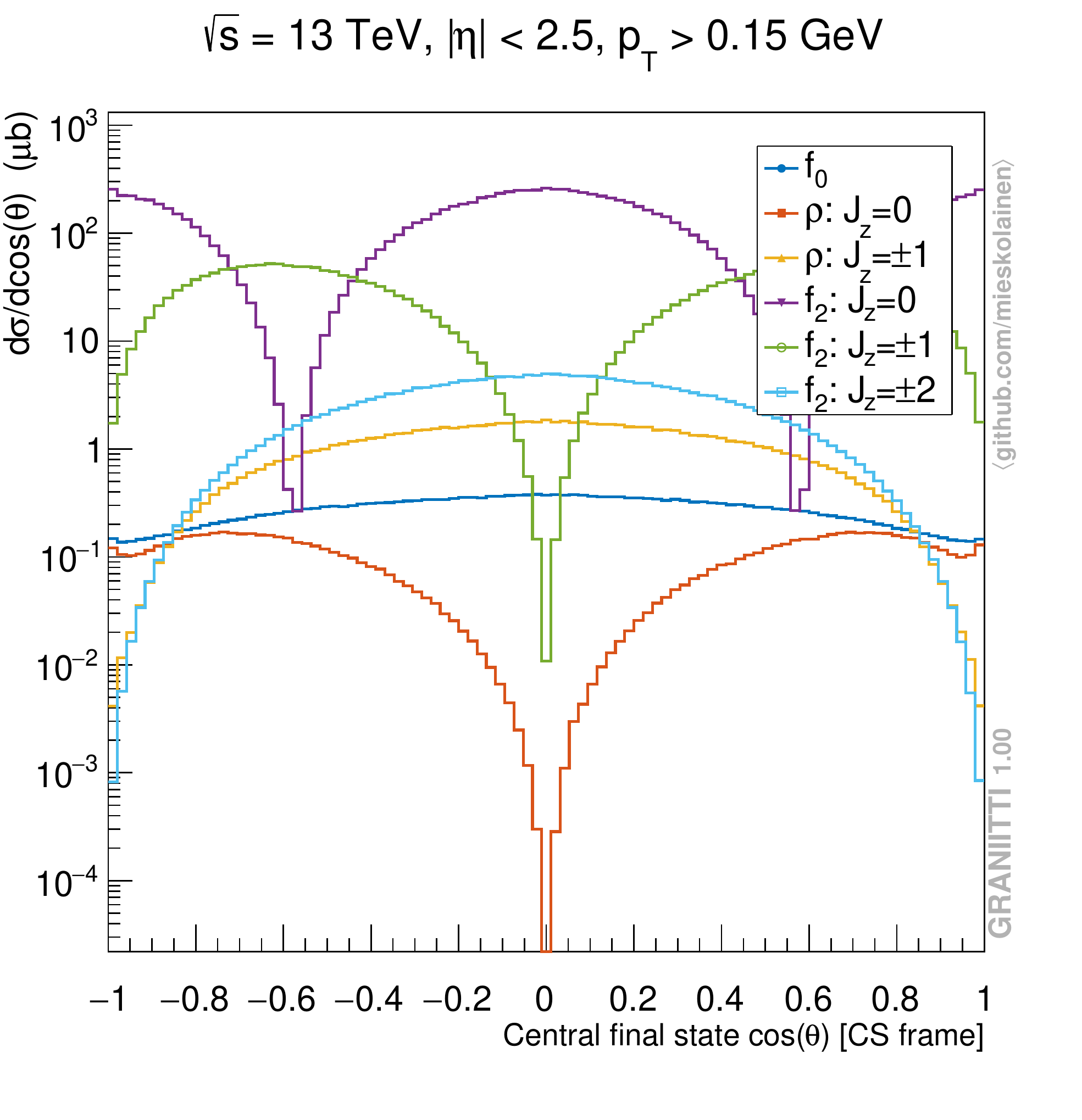}
\includegraphics[scale=0.3]{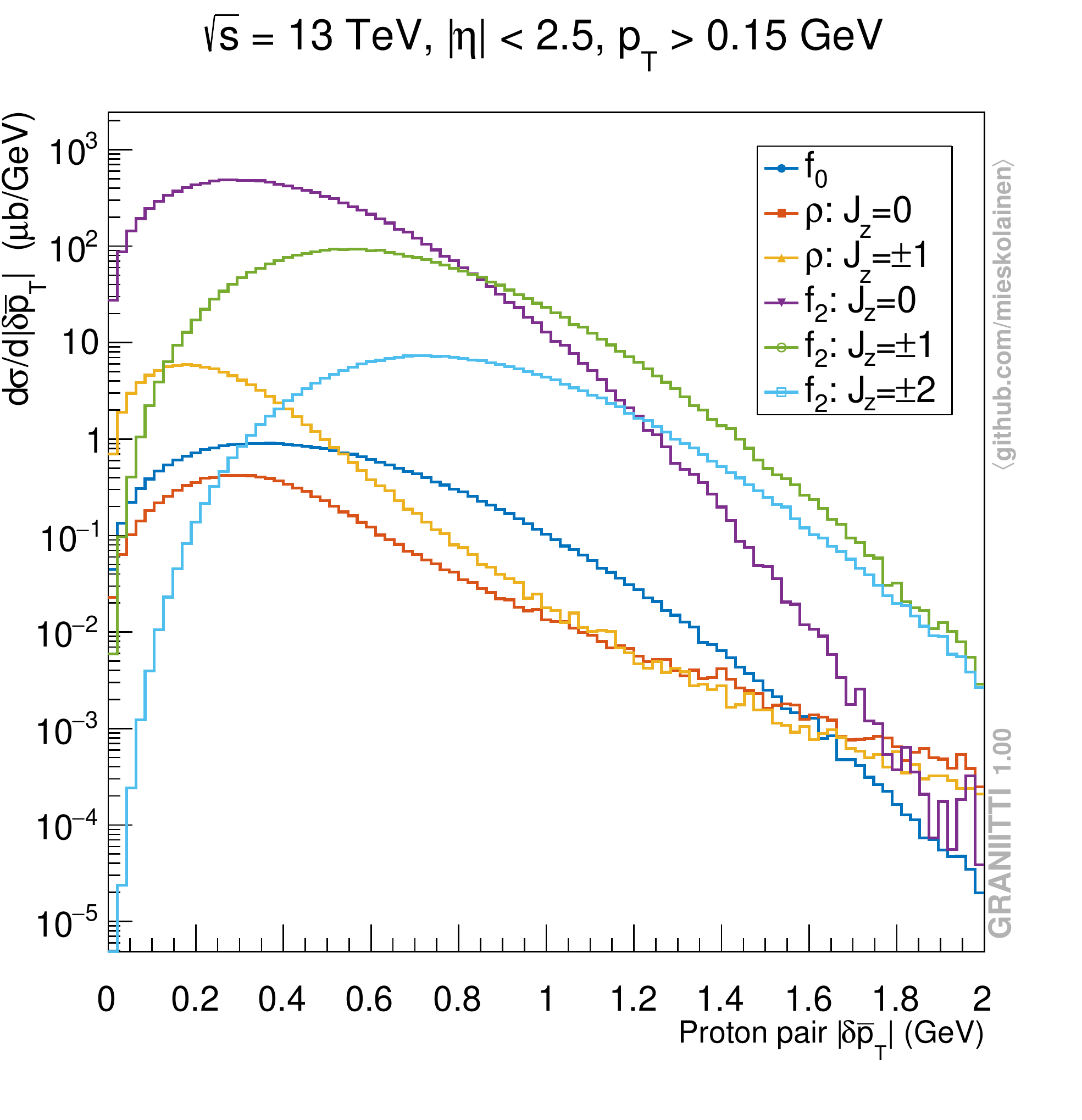}
\caption{Minimal Pomeron: Angular distribution $\cos(\theta)$ of $\pi^+$ in the Collins-Soper frame for different diagonal spin polarization components (left). Forward proton $p_T$-vector difference (right). For details, see~\cite{Mieskolainen:2019jpv}.}
\label{fig:minimal_pomeron}
\end{figure}

\begin{figure}
\centering
\includegraphics[scale=0.3]{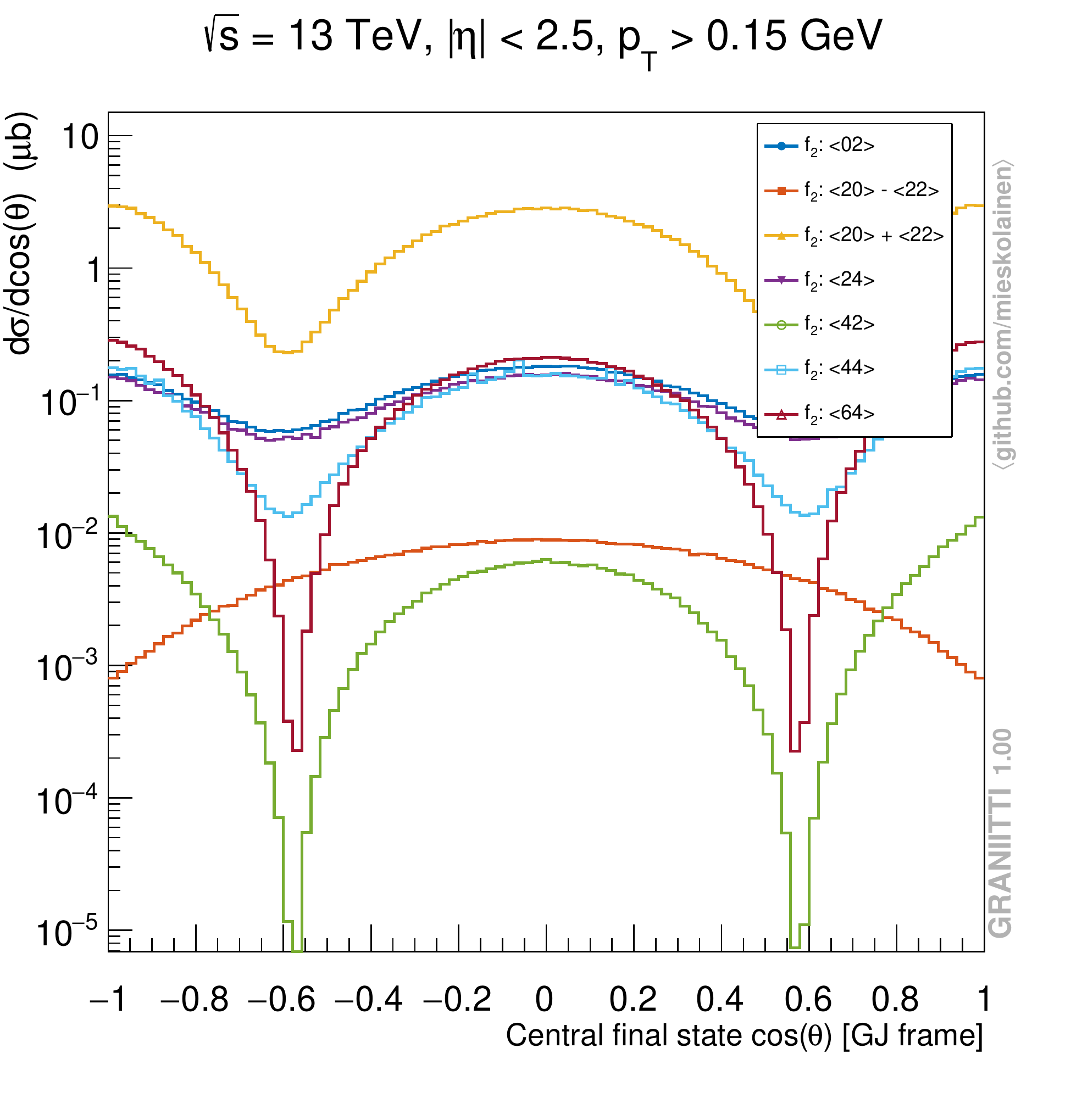}
\includegraphics[scale=0.3]{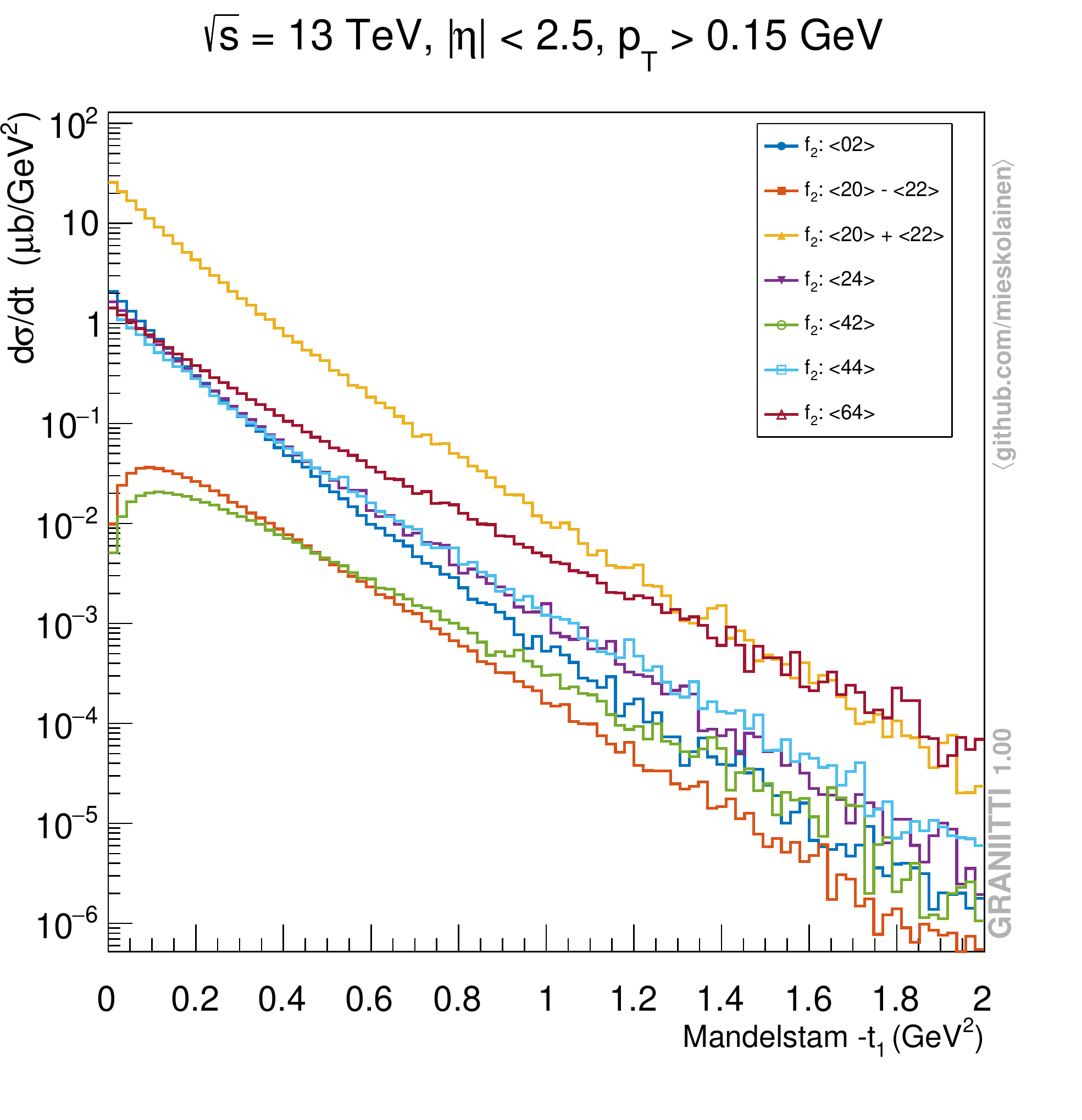}
\caption{Tensor Pomeron: Angular distribution $\cos(\theta)$ of $\pi^+$ in the Gottfried-Jackson frame (left) and Mandelstam $-t_1$ for 7 spin-2 couplings (right). For details, see~\cite{Mieskolainen:2019jpv}.}
\label{fig:tensor_pomeron}
\end{figure}

\begin{itemize}
\item[-] \textit{Minimal Pomeron}: a meson or baryon pair $2 \rightarrow 4$ continuum amplitudes~\cite{Harland-Lang:2013dia} together with Jacob-Wick helicity amplitudes for resonance decays and forward proton spin correlations, using numerical Wigner $3j$ algebra. For an illustration of distributions, see Figure~\ref{fig:minimal_pomeron}.
\item[-] \textit{Tensor Pomeron}: fully covariant $2 \rightarrow 4$ amplitudes~\cite{Lebiedowicz:2018sdt} for meson or baryon pairs both continuum and resonances, based on C++ template Lorentz algebra from numerical General Relativity and Dirac spinor algebra. For an illustration of distributions, see Figure~\ref{fig:tensor_pomeron}.
\end{itemize}

In addition to the soft QCD processes, the generator includes also $k_T$-EPA gamma flux driven gamma-gamma SM electroweak processes and the Durham model-based hard QCD scattering amplitudes together with Sudakov suppression integral factor and `Shuvaev transformed' gluon parton distributions based on numerical integral transforms of \textsc{Lhapdf6} gluon pdfs. \textsc{MadGraph} 5 amplitude import in \textsc{C++} format is supported for gamma-gamma processes. All processes can be generated either with elastic or inelastic forward protons and the differential screening (absorption) eikonal Pomeron loop amplitude turned on, which is crucial for cross-sections. For diagrammatic illustrations, see Figure~\ref{fig:diagrams}. In terms of kinematics and Monte Carlo, the full $2 \rightarrow N$ process is first constructed to be exact for $2 \rightarrow 3$ process which results in a lengthy polynomial expression due to three variable final masses, i.e. forward dissociation is allowed and importance sampled using \textsc{Vegas} together with custom analytic Jacobians. The central system phase space $1 \rightarrow N - 2$ is then further treated exactly with \textsc{Rambo}, utilizing exact recursive phase space factorization.

Beyond the set of readily available scattering amplitudes or matrix elements, arbitrary cascaded decay channels can be generated according to Breit-Wigner resonance weights $\times$ phase space with a decay chain interpreter and steering cards. Cascaded $1 \rightarrow 2$ resonance decays with arbitrary spin-parity combinations according to conservation laws are supported, where the spin dynamics and resulting correlations are computed using Jacob-Wick helicity amplitudes and user adjustable couplings. These cascaded decay chains allow probing different spin and parity hypothesis of the resonant central state, e.g. in decays of $X \rightarrow \rho^0 \rho^0 \rightarrow 2 \times (\pi^+ \pi^-)$.

\section{Comparisons with STAR data}
\label{sec:data}

\begin{figure}
\centering
\includegraphics[scale=0.6]{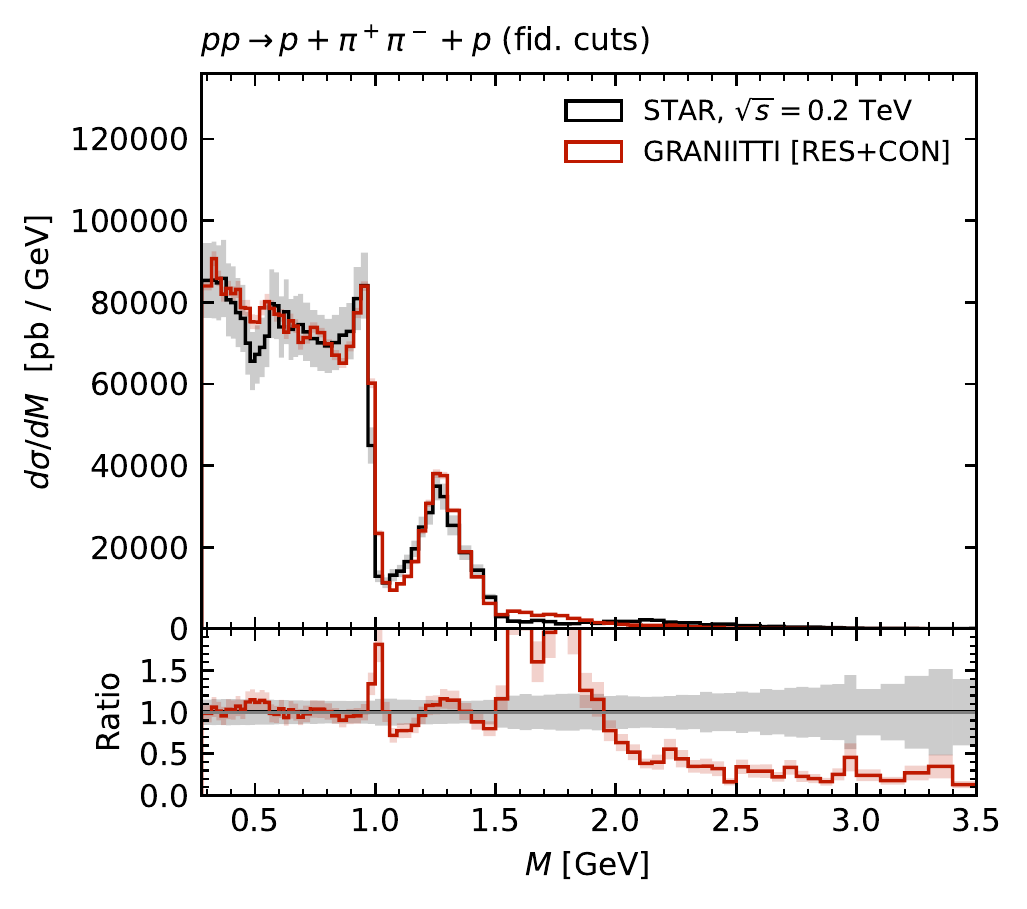}
\includegraphics[scale=0.6]{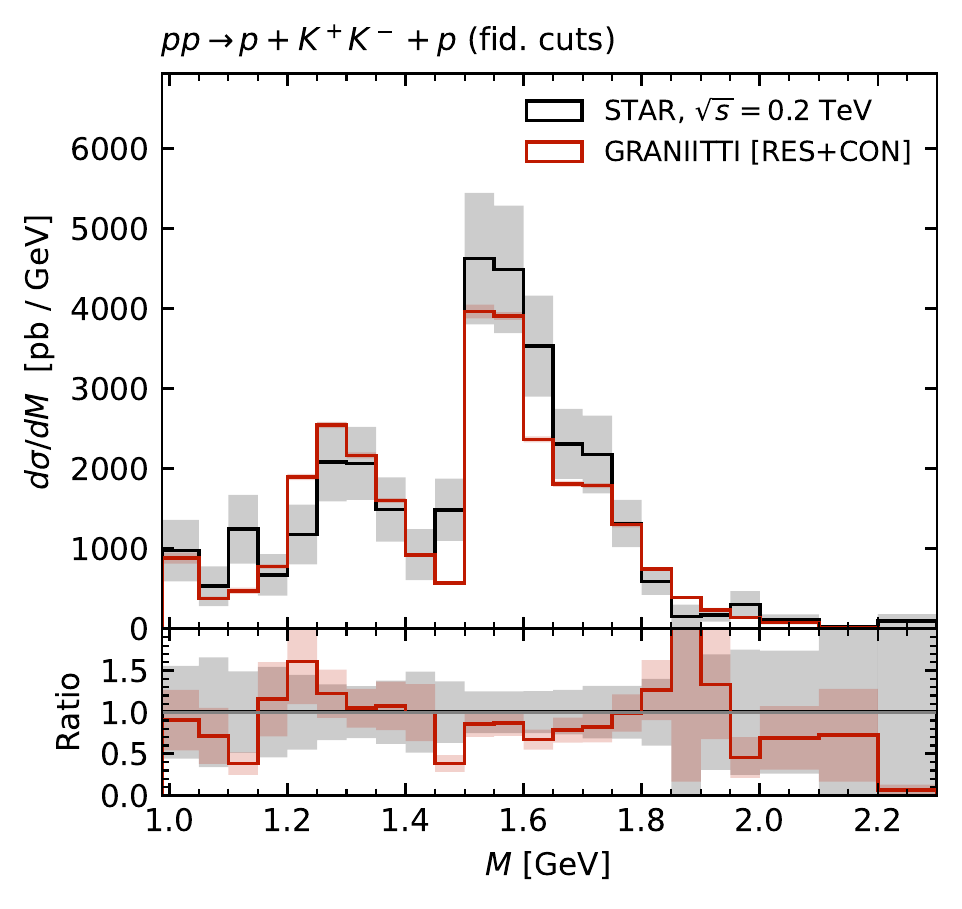}
\caption{The central system invariant mass in $\pi^+\pi^-$ (left) and $K^+K^-$ (right).}
\label{fig:masses}
\end{figure}

\begin{figure}
\centering
\includegraphics[scale=0.6]{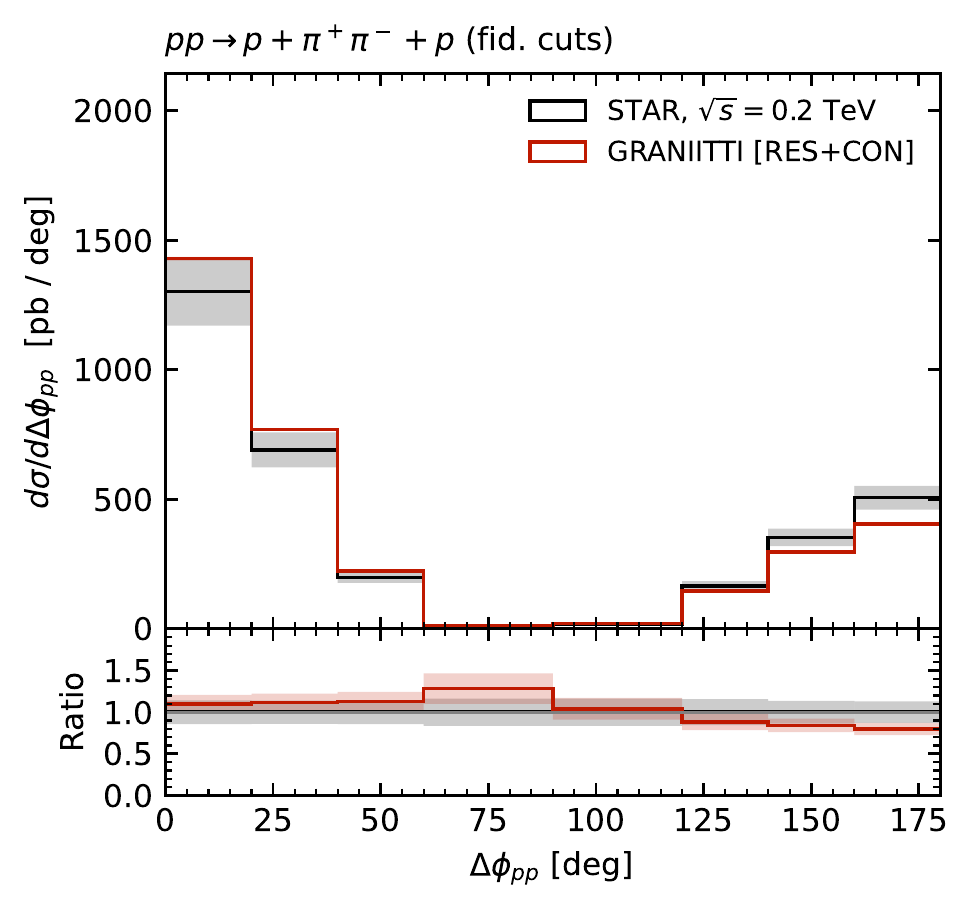}
\includegraphics[scale=0.6]{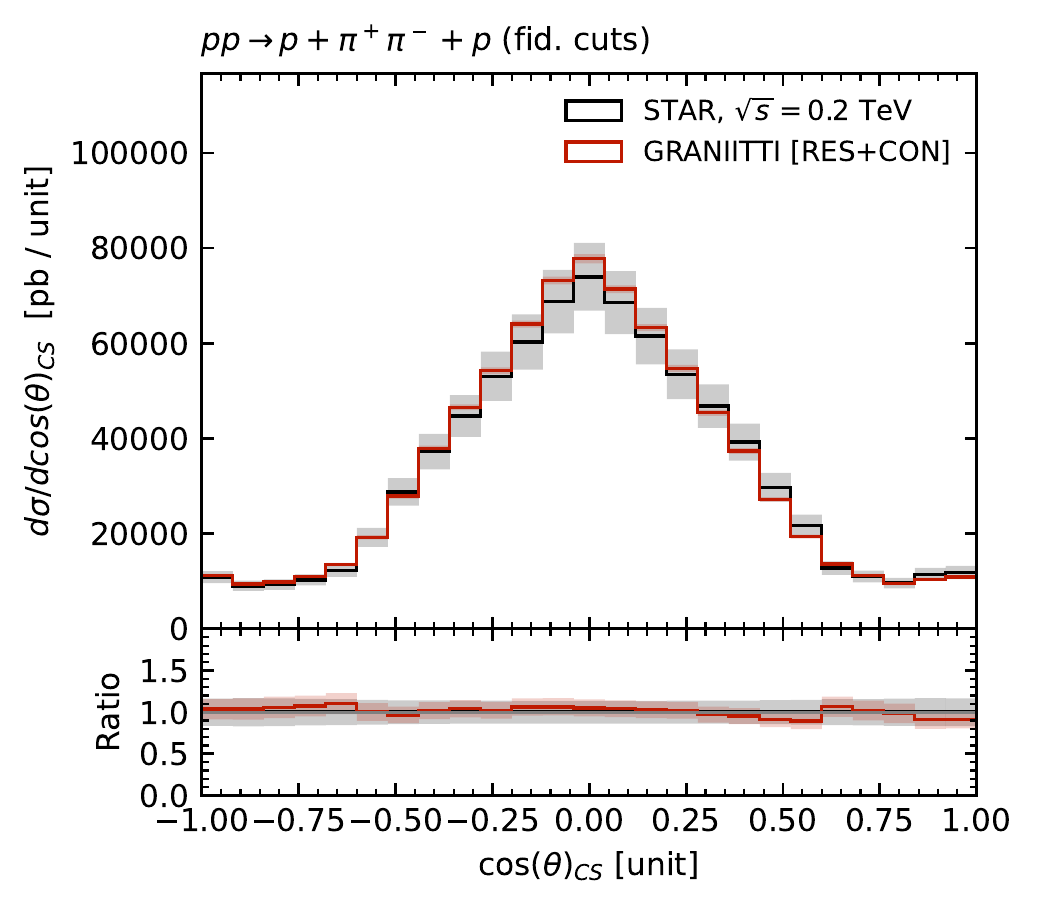}
\caption{Forward proton $\Delta \phi_{pp}$ (left) and CS-frame angle $\cos (\theta)$ of $\pi^+$ (right).}
\label{fig:forward}
\end{figure}

We compare the \textsc{Graniitti} minimal Pomeron model results with differential cross-section measurements of charged pion and kaon pairs with measured forward protons from the STAR experiment at RHIC~\cite{STAR:2020dzd}. These measurements include also proton-antiproton central pairs, which can be simulated. The fiducial cuts are as described for the most inclusive case in~\cite{STAR:2020dzd}. Data includes statistical, systematic and luminosity uncertainties summed in quadrature. A partial tune of the resonance couplings and continuum off-shell form factor has been done against the data here, leaving other soft parameters fixed such as the continuum couplings and eikonal Pomeron model. The simulations follow data in Figures~\ref{fig:masses} and \ref{fig:forward}, with an exception the high-mass pion pair tail, indicating need for further work on the screening loop effect vs the continuum amplitude (the form factor, parameters) and its possible perturbative descriptions and their matching. 

Interestingly, the data suggests for resonances $f_2(1270)$ and $f_2^{'}(1525)$ opposite spin-2 polarization of $J_z \simeq \pm 2$ and $J_z \simeq 0$ in the Collins-Soper (CS) frame, respectively. More differential angular distributions and e.g. spherical harmonics analysis~\cite{Mieskolainen:2019jpv} would be required for further conclusions in terms of glueball and hybrid quark-gluon state candidates.


\section{Computational technology towards deep learning}
\label{sec:technology}

\textsc{Graniitti} engine code is written in fully multithreaded C++17, enabling maximum CPU core utilization up to unlimited number of threads. Standard grid computing tools are naturally also supported, such as pre-computed integration arrays and different random seeds. Event output in the latest \textsc{HepMC3} format is provided. The crucial soft scattering amplitude parameter fitting or `tuning' challenge is accelerated using a novel approach. We have interfaced a cutting-edge \textsc{Raytune} library via Python steering code, which allows HPC-cluster distributable tuning of the event generator against Durham \textsc{HEPData} input. \textsc{Raytune} is used heavily in deep learning, especially in a gradient free neural network model hyperparameter tuning and reinforcement learning. The underlying global optimization algorithms are based on Bayesian optimization and evolutionary type schemes. Next, we outline ambitious steps towards the first deep learning-enhanced diffractive event generator.

First, understanding the proton structure fluctuations beyond integrated representations of proton form factors and parton distributions requires novel approaches for diffraction aka `deep Pomeron'. Generative deep learning techniques such as deep diffusion models based on non-equilibrium Langevin stochastics combined with deep neural networks provide a promising novel avenue to accomplish this goal, especially when the data-driven approach is combined with lattice field theory input.

Second, the Pomeron-Pomeron-resonance 3-point vertex of low-mass central production is highly non-perturbative and would be an excellent target to be learned directly from data using neural networks fitted against differential measurements, preferably event-by-event data such as CERN Open Data. Networks incorporating explicit Lorentz equivariance would be a natural choice for this task. Assuming the set of functions for resonances with different quark-gluon content would be universal, the learned neural vertices would allow predictive power via recycling the learned functions.

Third, to be able to accelerate the Monte Carlo integration and event generation efficiency, new methods beyond dimension-by-dimension factorized \textsc{Vegas} or phase space multichanneling are required. We have done preliminary studies in terms of invertible high-dimensional change of variable transformations, known as normalizing flows, which are chains of learned Jacobians with a fast analytic log-determinant. The results are promising, but the challenge is in fully generic solutions which may require a new type of flow layers due to challenging multiparticle Lorentz manifolds.

\bibliography{refs} 

\begin{thebibliography}{1}

\bibitem{Mieskolainen:2019jpv}
M.~Mieskolainen.
\newblock {GRANIITTI: A Monte Carlo Event Generator for High Energy
  Diffraction}.
\newblock 2019.
\newblock arXiv:1910.06300.

\bibitem{Harland-Lang:2013dia}
L.~A. Harland-Lang, V.~A. Khoze, and M.~G. Ryskin.
\newblock {Modelling exclusive meson pair production at hadron colliders}.
\newblock {\em Eur. Phys. J. C}, 74, 2014.

\bibitem{Lebiedowicz:2018sdt}
P.~Lebiedowicz, O.~Nachtmann, and A.~Szczurek.
\newblock {Central exclusive diffractive production of $p \bar{p}$ pairs in
  proton-proton collisions at high energies}.
\newblock {\em Phys. Rev. D}, 97(9), 2018.

\bibitem{STAR:2020dzd}
J.~Adam et~al.
\newblock {Measurement of the central exclusive production of charged particle
  pairs in proton-proton collisions at $\sqrt{s} = 200$ GeV with the STAR
  detector at RHIC}.
\newblock {\em JHEP}, 07(07), 2020.

\end{thebibliography}

\end{document}